\documentclass[a4paper,fleqn,usenatbib]{mnras}

\usepackage{newtxtext,newtxmath}

\usepackage[T1]{fontenc}
\usepackage{ae,aecompl}

\usepackage{graphicx}	
\usepackage{amsmath}	
\usepackage{amssymb}	

\newcommand{\be}{\begin{eqnarray}}
\newcommand{\ee}{\end{eqnarray}}

\title[Thermal spectra of thin disks of finite thickness]{Thermal spectra of thin accretion disks of finite thickness around Kerr black holes}

\author[Zhou et al.]{
Menglei~Zhou,$^{1}$ 
Askar~B.~Abdikamalov,$^{1}$
Dimitry~Ayzenberg,$^{1}$
Cosimo~Bambi,$^{1}$\thanks{Corresponding author: bambi@fudan.edu.cn} \newauthor
Victoria~Grinberg,$^{2}$
and Ashutosh~Tripathi$^{1}$
\\
$^{1}$Center for Field Theory and Particle Physics and Department of Physics, Fudan University, 200438 Shanghai, China\\
$^{2}$Institut f\"ur Astronomie und Astrophysik (IAAT), Eberhard-Karls Universit\"at T\"ubingen, 72076 T\"ubingen, Germany
}

\begin{document}
\label{firstpage}
\pagerange{\pageref{firstpage}--\pageref{lastpage}}
\maketitle

\begin{abstract}
The analysis of the thermal spectrum of geometrically thin and optically thick accretion disks of black holes, the so-called continuum-fitting method, is one of the leading techniques for measuring black hole spins. Current models normally approximate the disk as infinitesimally thin, while in reality the disk thickness is finite and increases as the black hole mass accretion rate increases. Here we present an XSPEC model to calculate the multi-temperature blackbody spectrum of a thin accretion disk of finite thickness around a Kerr black hole. We test our new model with an \textsl{RXTE} observation of the black hole binary GRS~1915+105. We find that the spin value inferred with the new model is slightly higher than the spin value obtained with a model with an infinitesimally thin disk, but the difference is small and the effect is currently subdominant with respect to other sources of uncertainties in the final spin measurement.
\end{abstract}




\section{Introduction}

Black holes are among the most exotic objects that can be found in the contemporary Universe~\citep{2018AnP...53000430B,2019arXiv190603871B}. According to Einstein's theory of general relativity, a black hole should be completely characterized by its mass $M$, spin angular momentum $J$, and electric charge $Q$~\citep{1971PhRvL..26..331C,1975PhRvL..34..905R}, but the latter is thought to be completely negligible for astrophysical black holes. The mass $M$ is relatively easy to measure, by studying the orbital motion of individual stars orbiting the black hole~\citep[see, for instance,][]{2014SSRv..183..223C,2008ApJ...689.1044G}. The measurement of the spin is definitively more challenging. The spin of a rotating object has no gravitational effects in Newtonian gravity, so black hole spin measurements require the analysis of relativistic phenomena occurring in the strong gravity region of the black hole.

There are currently two leading techniques for measuring the spin of accreting black holes: the continuum-fitting method~\citep{1997ApJ...482L.155Z,2011CQGra..28k4009M,2014SSRv..183..295M} and X-ray reflection spectroscopy~\citep{2006ApJ...652.1028B,2014SSRv..183..277R}. The continuum-fitting method is the analysis of the thermal spectrum of geometrically thin and optically thick accretion disks of black holes and is normally used for stellar-mass black holes only. Indeed, the temperature of a thin accretion disk scales as $M^{-0.25}$ and the disk mainly emits in the soft X-ray band for stellar-mass black holes and in the optical/UV bands for supermassive black holes. In the latter case, dust absorption limits the capability of an accurate measurement of the thermal spectrum and, in turn, the possibility of measuring the black hole spin. X-ray reflection spectroscopy refers to the analysis of the reflection spectrum of thin accretion disks and can measure the spin of black holes of any mass.

The standard framework to describe geometrically thin and optically thick accretion disks of black holes is the Novikov-Thorne model~\citep{1973blho.conf..343N,1974ApJ...191..499P}, which is normally thought to be a good approximation for accretion disks of sources in the thermal state and with an Eddington-scaled accretion luminosity between a few percent to about 30\%~\citep{2006ApJ...652..518M,2010ApJ...718L.117S,2010MNRAS.408..752P,2011MNRAS.414.1183K}. Common models for the continuum-fitting method and X-ray reflection spectroscopy employ the Novikov-Thorne model and approximate the disk as infinitesimally thin, with the particles of the gas moving on nearly-geodesic, equatorial, circular orbits. However, in reality the disk has a finite thickness, which increases as the mass accretion rate increases.

In this paper, we present a model to calculate the thermal spectrum of a geometrically thin and optically thick accretion disk of finite thickness around Kerr black holes. We implement the disk model proposed in \citet{2018ApJ...855..120T} in the multi-temperature blackbody model {\sc nkbb}~\citep{2019PhRvD..99j4031Z,2020arXiv200108391T}. A ray-tracing code calculates the transfer function of the spacetime for a disk with finite thickness~\citep{1975ApJ...202..788C} and the transfer functions for a grid of black hole spins, mass accretion rates, and disk inclination angles are stored in a FITS file. The model {\sc nkbb} can be used in XSPEC~\citep{xspec} and reads the FITS file of the transfer functions during the data analysis.

To illustrate the impact of the disk thickness on the spin measurement, we analyze an \textsl{RXTE} observation of the black hole binary GRS~1915+105 with {\sc nkbb}, either assuming an infinitesimally thin accretion disk and employing the new version of the model with a disk of finite thickness. We find that the impact of the disk thickness on the estimate of the spin of the black hole in GRS~1915+105 is small. The value of the black hole spin inferred with the new model is slightly higher than the value found with the model assuming an infinitesimally thin disk. For the quality of the data analyzed, as well as considering the current typical uncertainties of black hole masses, distances, and inclination angles, the correction on the black hole spin measurement from the disk thickness can be ignored, but in the future, with more accurate and precise spin measurements, it may become necessary to take it into account for very high disk inclination angles.

The content of the paper is as follows. In Section~\ref{s-ad}, we review the accretion disk models with infinitesimally thin disk and with disk of finite thickness. In Section~\ref{s-m}, we describe the construction of the model. In Section~\ref{s-grs}, we analyze an \textsl{RXTE} observation of the black hole binary GRS~1915+105 and we compare the black hole spin measurements obtained, respectively, with a model with infinitesimally thin disk and a model with a disk of finite thickness. We discuss our results in Section~\ref{s-d}.


\section{Accretion disk models}\label{s-ad}

The Novikov-Thorne model is the standard framework for the description of geometrically thin and optically thick accretion disks around black holes~\citep{1973blho.conf..343N,1974ApJ...191..499P}. In this paper, we will assume that the spacetime metric is described by the Kerr solution~\citep{1963PhRvL..11..237K}, but the considerations and the expressions reported in this section hold for any stationary and axisymmetric black hole spacetime with a line element in spherical-like coordinates $(t,r,\theta,\phi)$ that can be written as\footnote{Note that this is not the most general line element for a stationary and axisymmetric spacetime; in general, $g_{tr}$ may also be non-vanishing. Nevertheless, black hole solutions in general relativity and in many other theories of gravity have $g_{tr} = 0$.}
\be
ds^2 = g_{tt} dt^2 + 2 g_{t\phi} dt d\phi + g_{rr} dr^2 + g_{\theta\theta} d\theta^2 + g_{\phi\phi} d\phi^2 \, ,
\ee 
where all the metric coefficients are independent of $t$ and $\phi$. Note that in this section we assume a metric with signature $(-+++)$ and employ units in which $c=1$. The details of the calculations can be found in \citet{2017bhlt.book.....B,2012ApJ...761..174B}.

If we approximate the disk as infinitesimally thin, the particles of the fluid move on nearly geodesic, circular ($r= {\rm constant}$), equatorial ($\theta=\pi/2$) orbits. We write the geodesic equations in the form
\be
\frac{d}{d\tau} \left(g_{\mu\nu} \dot{x}^\nu \right) 
= \frac{1}{2} \left(\partial_\mu g_{\nu\rho}\right) \dot{x}^\nu \dot{x}^\rho \, .
\ee
Since the particles of the fluid have $\dot{r} = \ddot{r} = \dot{\theta} = 0$, for $\mu=r$ we have
\be\label{eq-omega0}
\left(\partial_r g_{tt}\right) \dot{t}^2
+ 2 \left(\partial_r g_{t\phi}\right) \dot{t} \dot{\phi}
+ \left(\partial_r g_{\phi\phi}\right) \dot{\phi}^2 = 0 \, .
\ee 
The angular velocity of the fluid as measured by an observer at infinity is $\Omega = \dot{\phi}/\dot{t}$. From Eq.~(\ref{eq-omega0}) we find
\be\label{eq-omega}
\Omega_\pm = \frac{- \left(\partial_r g_{t\phi}\right) \pm 
\sqrt{\left(\partial_r g_{t\phi}\right)^2 - \left(\partial_r g_{tt}\right)
\left(\partial_r g_{\phi\phi}\right)}}{\partial_r g_{\phi\phi}} \, ,
\ee
where the upper (lower) sign refers to an accretion disk with angular momentum parallel (antiparallel) to the black hole spin.

From the conservation of the rest-mass of the particles of the fluid $g_{\mu\nu} u^\mu u^\nu = - 1$ and the conditions $\dot{r} = \dot{\theta} = 0$ on the fluid motion, we derive $\dot{t}$
\be\label{eq-t}
\dot{t} = \frac{1}{\sqrt{- g_{tt} - 2 \Omega g_{t\phi} - \Omega^2 g_{\phi\phi}}} \, .
\ee
As the disk is infinitesimally thin, its surface is on the equatorial plane, and the 4-velocity of the particles on the surface of the disk is $u^\mu = (1,0,0,\Omega) \, \dot{t}$, where the expressions of $\dot{t}$ and $\Omega$ are, respectively, in Eq.~(\ref{eq-t}) and Eq.~(\ref{eq-omega}), and in both cases all quantities are evaluated on the equatorial plane ($\theta=\pi/2$).

\begin{figure*}
\begin{center}
\includegraphics[width=0.33\textwidth,trim={1.0cm 0cm 1.5cm 0cm},clip]{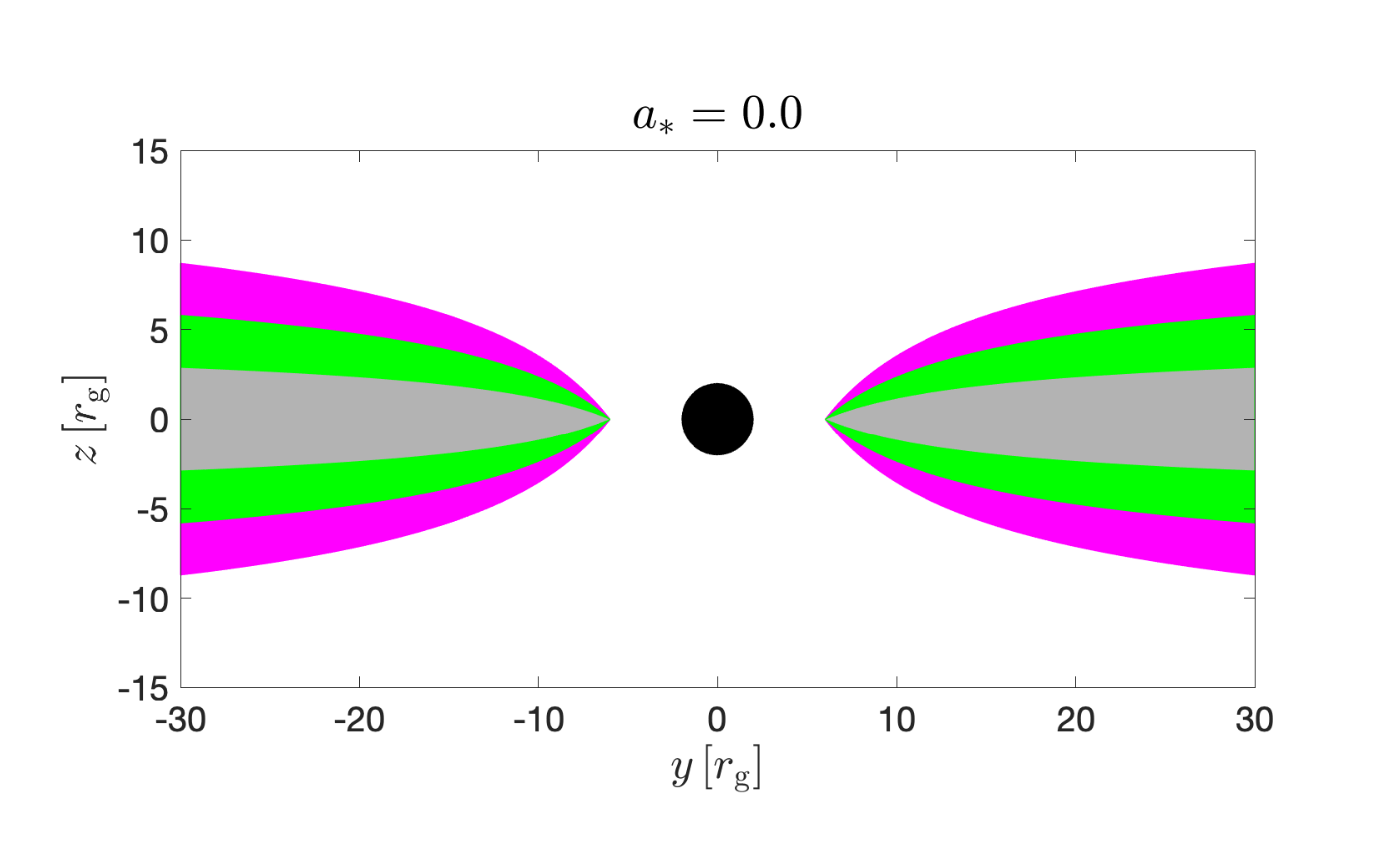}
\includegraphics[width=0.33\textwidth,trim={1.0cm 0cm 1.5cm 0cm},clip]{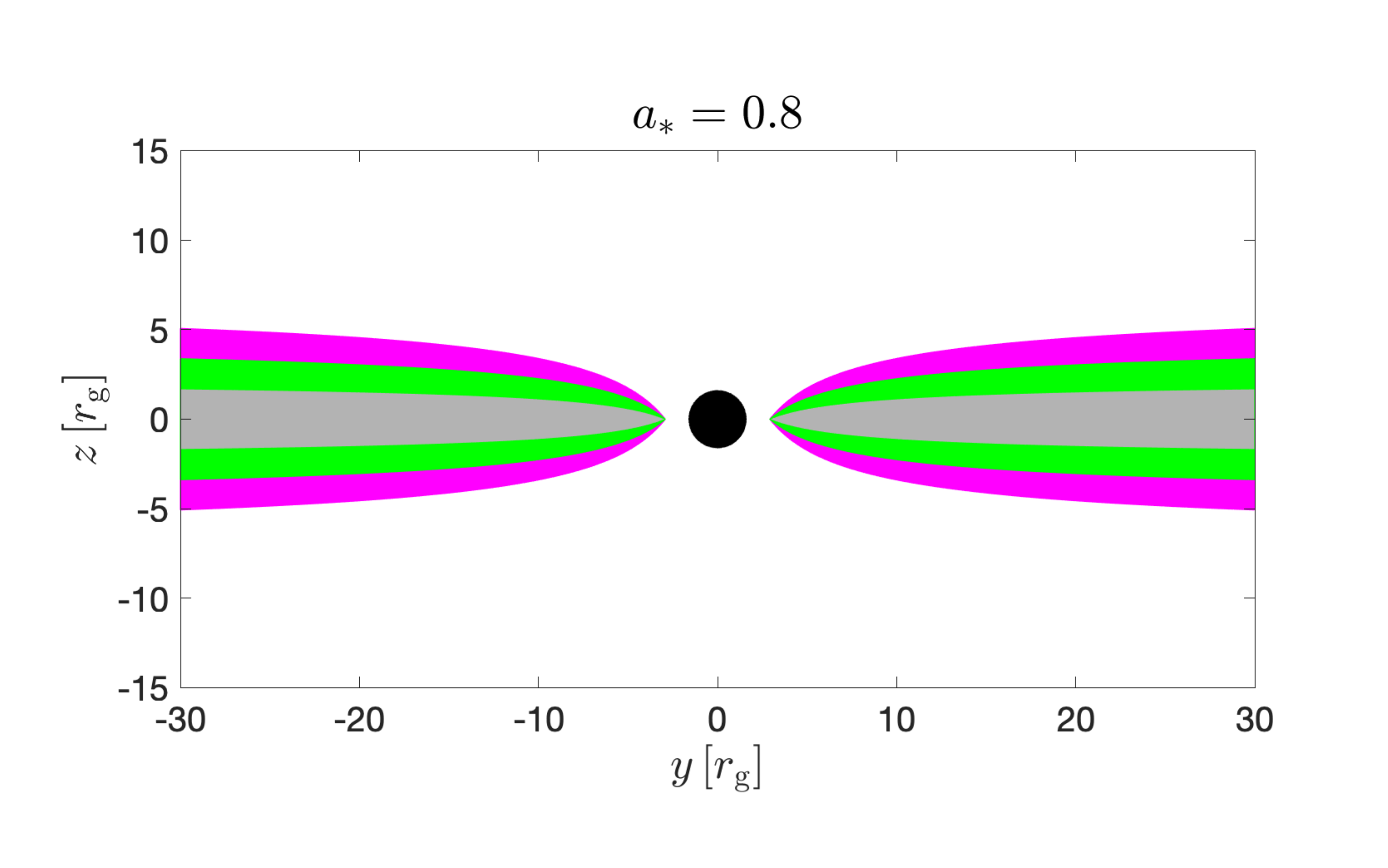}
\includegraphics[width=0.33\textwidth,trim={1.0cm 0cm 1.5cm 0cm},clip]{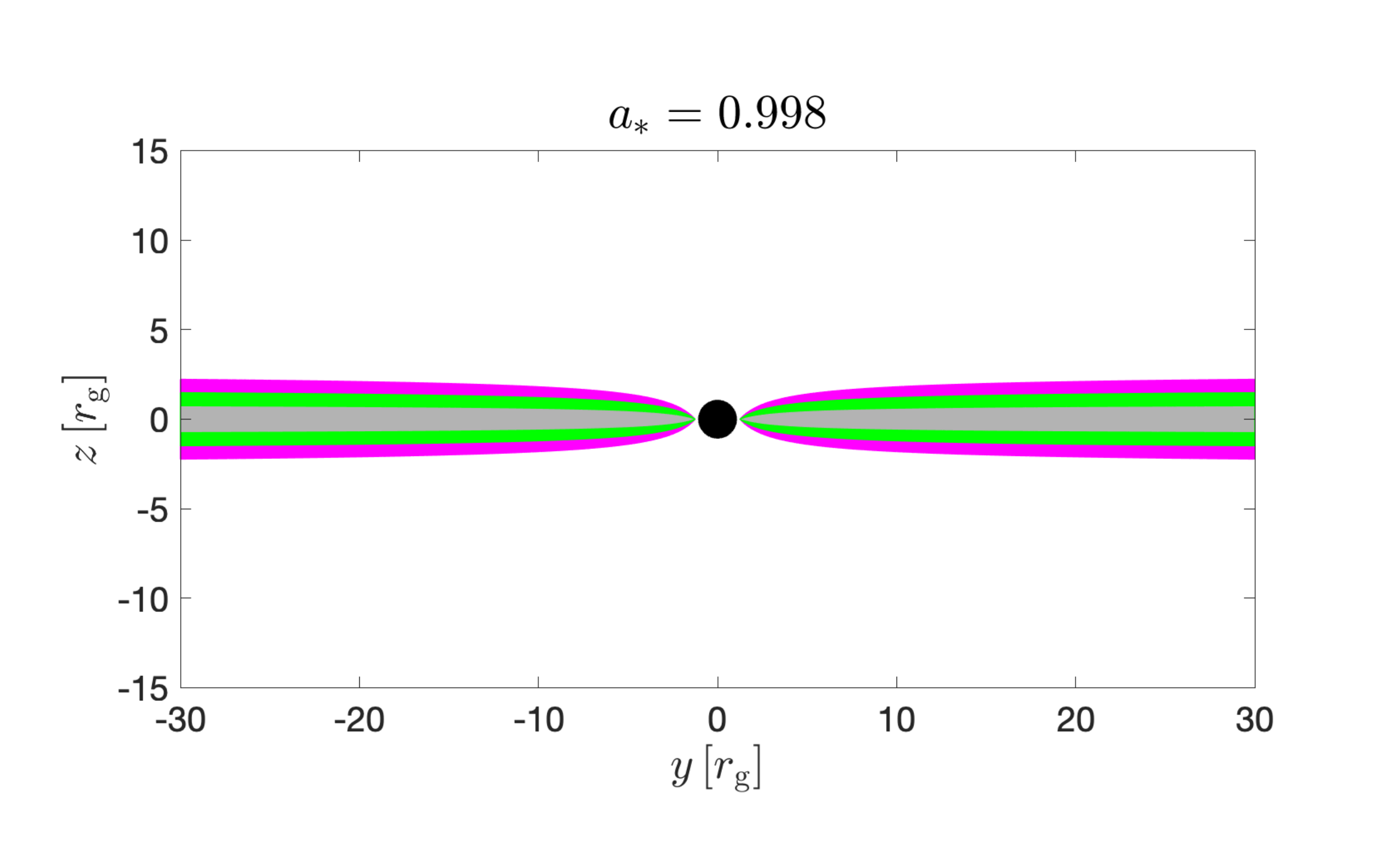} 
\end{center}
\vspace{-0.4cm}
\caption{Examples of accretion disk profiles, following~\citet{2018ApJ...868..109T}. 
The black hole spin parameter is $a_* = 0$ (left panel), 0.8 (central panel), and 0.998 (right panel). The mass accretion rate is $\dot{M}/\dot{M}_{\rm Edd} = 0.1$ (gray), 0.2 (green), and 0.3 (magenta). $y$- and $z$-axes in units of the gravitational radius $r_{\rm g} = M$. See the text for more details. \label{fig:disks}}
\end{figure*}

\citet{2018ApJ...855..120T} have recently proposed a simple framework to take the thickness of the disk into account. The mid-plane of the accretion disk is still on the equatorial plane $\theta = \pi/2$. For a radiatively dominated, optically thick disk, the pressure scale height is~\citep{1973A&A....24..337S}
\be
H = \frac{3}{2} \frac{1}{\eta} \left(\frac{\dot{M}}{\dot{M}_{\rm Edd}}\right) 
\left( 1 - \sqrt{\frac{R_{\rm ISCO}}{\rho}} \right) \, ,
\ee
where $\eta = 1 - E_{\rm ISCO}$ the radiative efficiency of the Novikov-Thorne accretion disk, $E_{\rm ISCO}$ is the specific energy of a test-particle at the radius of the innermost stable circular orbit (ISCO) on the equatorial plane, $\dot{M}/\dot{M}_{\rm Edd}$ is the Eddington-scaled mass accretion rate, $R_{\rm ISCO}$ is the ISCO radius, and $\rho = r \sin\theta$ is the pseudo-cylindrical radius. The surface of the disk is set at
\be\label{eq-z}
z(\rho) = 2 H (\rho) \, 
\ee
All the particles of the fluid with the same pseudo-cylindrical radius $\rho$ are supposed to rotate with the angular velocity of a test-particle on a geodesic, equatorial, circular orbit at that value of $\rho$. Since $\eta$ and $R_{\rm ISCO}/M$ depend on the black hole spin parameter $a_*$ in the Kerr spacetime, black holes with the same Eddington-scaled mass accretion rate can have disks with different thickness according to the value of their spin parameter. Fig.~\ref{fig:disks} shows the disk profiles for a central black hole with $a_* = 0$, 0.8, and 0.998, and an Eddington-scaled mass accretion $\dot{M}/\dot{M}_{\rm Edd} = 0.1$, 0.2, and 0.3. For a given $a_*$, $\dot{M}/\dot{M}_{\rm Edd}$ is the parameter regulating the thickness of the disk.

In the model proposed in \citet{2018ApJ...855..120T}, the surface of the disk is determined by Eq.~(\ref{eq-z}). The 4-velocity of the particles of the fluid on the surface of the disk is still $u^\mu = (1,0,0,\Omega) \, \dot{t}$, with $\Omega$ still evaluated on the equatorial plane (i.e. $r = \rho$ and $\theta = \pi/2$) and $\dot{t}$ is evaluated from Eq.~(\ref{eq-t}) with $\Omega$ evaluated on the equatorial plane and the metric coefficients evaluated at the exact point on the surface of the disk.

We note that such a simple model only adds the finite thickness of the disk. The most significant missing physical process is the advection of energy by the radial inflow, which would slightly change, and ultimately decrease, the energy radiated from the disk surface.

\section{Thermal spectra of accretion disks of finite tickness}\label{s-m}

The finite thickness disk geometry can be implemented in the relativistic thermal model {\sc nkbb}~\citep{2019PhRvD..99j4031Z,2020arXiv200108391T}. While {\sc nkbb} is specifically designed for testing the Kerr metric~\citep{2017RvMP...89b5001B}, in this work we will ignore such a possibility and we will only consider the Kerr background, either with an infinitesimally thin disk or a disk of finite thickness. The extension to non-Kerr backgrounds is not so straightforward, because the FITS file is now for $(a_*, \dot{M}/\dot{M}_{\rm Edd},i)$, where $\dot{M}/\dot{M}_{\rm Edd}$ replaces the deformation parameter of the spacetime of the normal version of {\sc nkbb} and a FITS file with 4~parameters would become too heavy.

The model employs the formalism of the transfer function proposed by Cunningham~\citep{1975ApJ...202..788C,1995CoPhC..88..109S}. We consider a static observer at spatial infinity. The flux of the accretion disk as measured by the distant observer can be written as
\be
F_{\rm o} (\nu_{\rm o}) = \frac{1}{D^2} \int I_{\rm o} (\nu_{\rm o}) \, dXdY 
=  \frac{1}{D^2} \int g^3 I_{\rm e} (\nu_{\rm e}) \, dXdY \, ,
\ee
where $\nu_{\rm o}$ and $\nu_{\rm e}$ are the photon frequencies in the rest-frame of the distant observer and of the gas, respectively, $X$ and $Y$ are the Cartesian coordinates of the plane of the distant observer, $D$ is the distance of the observer from the source, $I_{\rm o}$ and $I_{\rm e}$ are the specific intensities of the radiation in the rest-frame of the distant observer and of the gas, respectively, $I_{\rm o} = g^3 I_{\rm e}$ follows from Liouville's theorem~\citep{1966AnPhy..37..487L}, and $g$ is the redshift factor
\be
g = \frac{\nu_{\rm o}}{\nu_{\rm e}} = \frac{(k_\mu)_{\rm o} u^\mu_{\rm o}}{(k_\nu)_{\rm e} u^\nu_{\rm e}} \, ,
\ee
where $u^\mu_{\rm o} = (1,0,0,0)$ is the 4-velocity of the distant observer, $u^\mu_{\rm e} = (1,0,0,\Omega) \, \dot{t}$ is the 4-velocity of the gas on the surface of the accretion disk (which changes if we assume infinitesimally thin disk or finite thickness disk), $k^\mu$ is the 4-momentum of the photon, which is evaluated, respectively, at the detection point in the numerator and at the emission point in the denominator.

Introducing the transfer function $f$, the flux of the accretion disk can be written as~\citep{1975ApJ...202..788C}
\be\label{eq-F}
F_{\rm o} (\nu_{\rm o}) = \frac{1}{D^2} \int_{R_{\rm in}}^{R_{\rm out}} \int_0^1 
\frac{\pi r_{\rm e} g^2 f(g^*,r_{\rm e},i)}{\sqrt{g^* (1 - g^*)}} \, I_{\rm e} \, dg^* \, dr_{\rm e} 
\ee
where $R_{\rm in}$ and $R_{\rm out}$ are the inner and the outer edge of the accretion disk, respectively, and $g^*$ is the relative redshift factor defined by
\be
g^* = \frac{g - g_{\rm min}}{g_{\rm max} - g_{\rm min}} \, ,
\ee
where $g_{\rm min} = g_{\rm min} (r_{\rm e},i)$ and $g_{\rm max} = g_{\rm max} (r_{\rm e},i)$ are, respectively, the minimum and the maximum values of the redshift factor $g$ for the photons emitted from the radial coordinate $r_{\rm e}$ and for an inclination angle of the disk $i$ (i.e., the angle between the black hole spin and the line of sight of the distant observer). $f(g^*,r_{\rm e},i)$ is the transfer function 
\be
f(g^*,r_{\rm e},i) = \frac{g \sqrt{g^* (1 - g^*)}}{\pi r_{\rm e}} 
\left| \frac{\partial \left(X,Y\right)}{\partial \left(g^*,r_{\rm e}\right)} \right| \, ,
\ee
where $|\partial \left(X,Y\right)/\partial \left(g^*,r_{\rm e}\right)|$ is the Jacobian between the Cartesian coordinates of the screen of the distant observer and the disk variables $g^*$ and $r_{\rm e}$.

\begin{figure}
\begin{center}
\includegraphics[width=0.45\textwidth,trim={0cm 0cm 0cm 0cm},clip]{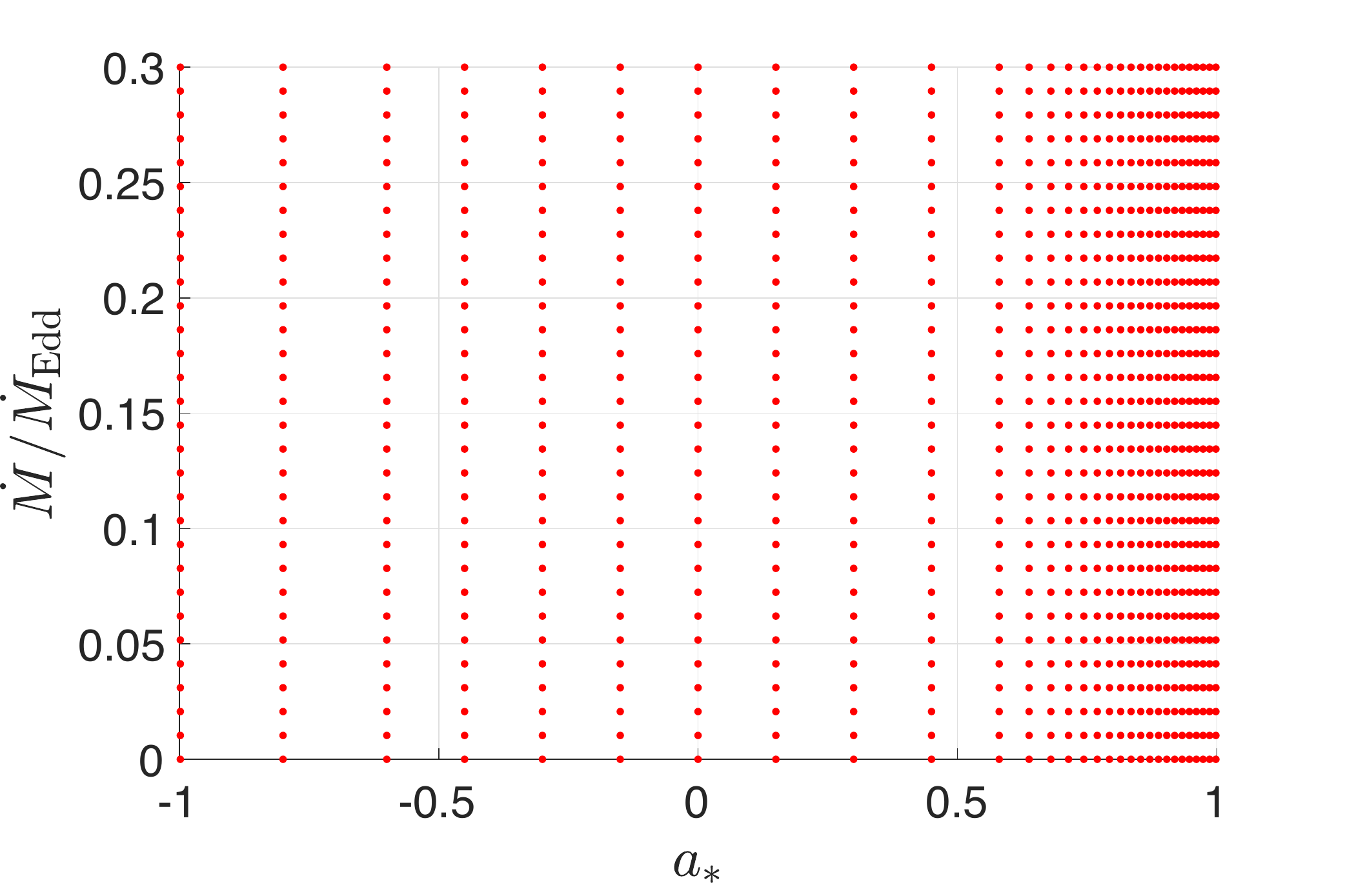}
\end{center}
\vspace{-0.2cm}
\caption{Grid points of the FITS file of the transfer function on the plane spin parameter $a_*$ vs mass accretion rate $\dot{M}$. Note that the grid spacings are non-uniform in $a_*$ and uniform in $\dot{M}$. \label{fig:grid}}
\end{figure}

\begin{figure*}
\begin{center}
\includegraphics[width=0.3\textwidth]{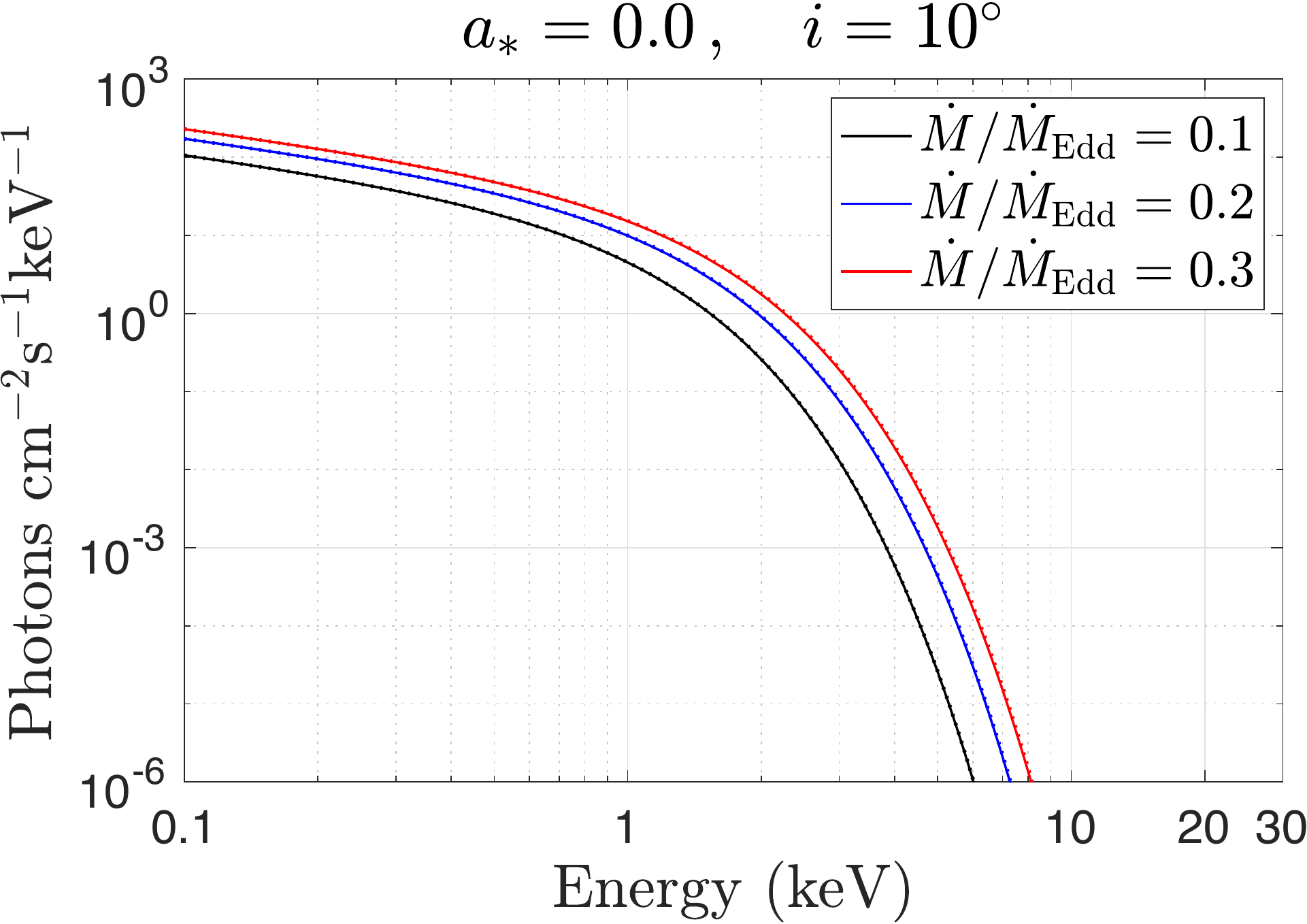} \hspace{0.4cm}
\includegraphics[width=0.3\textwidth]{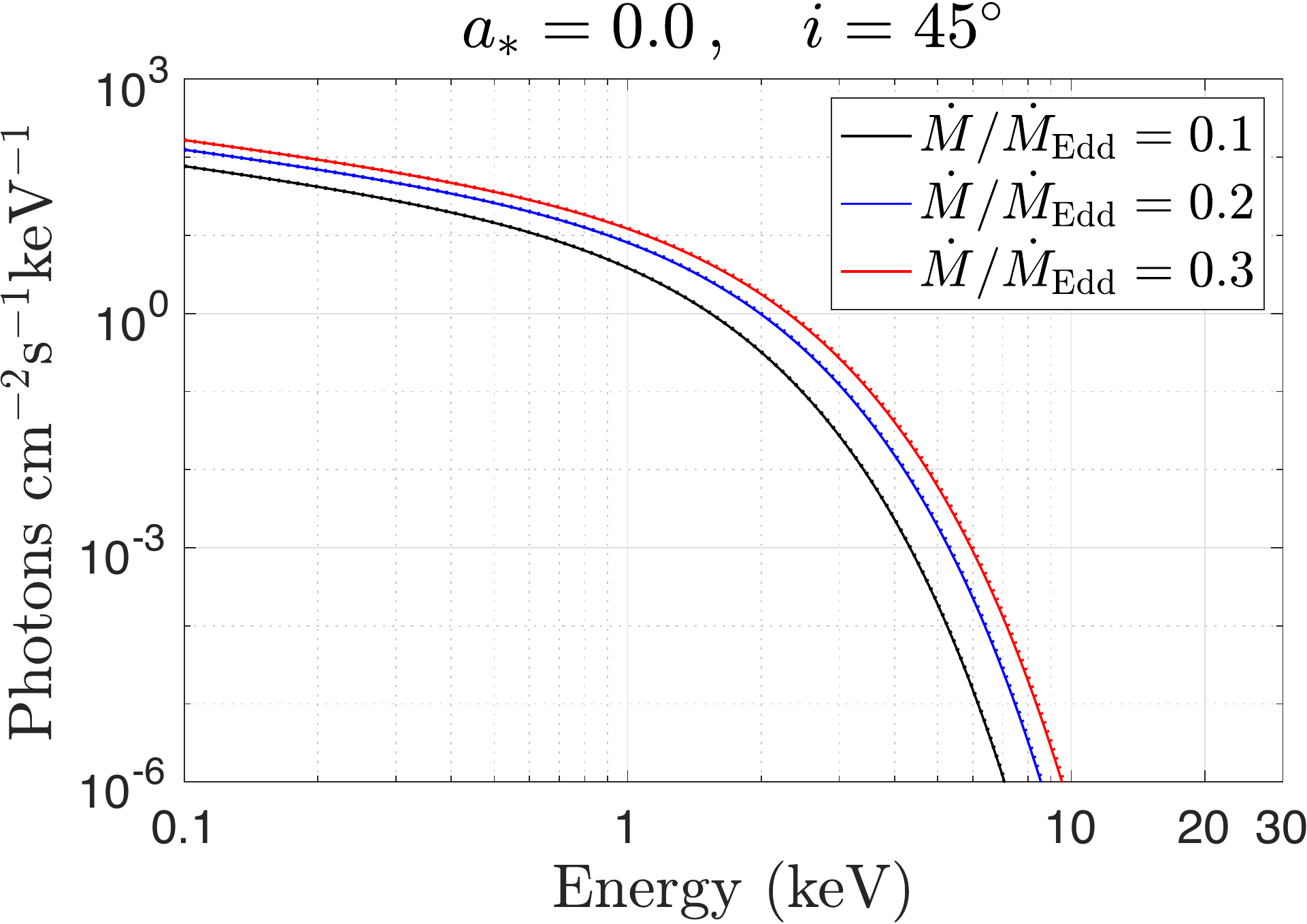} \hspace{0.4cm}
\includegraphics[width=0.3\textwidth]{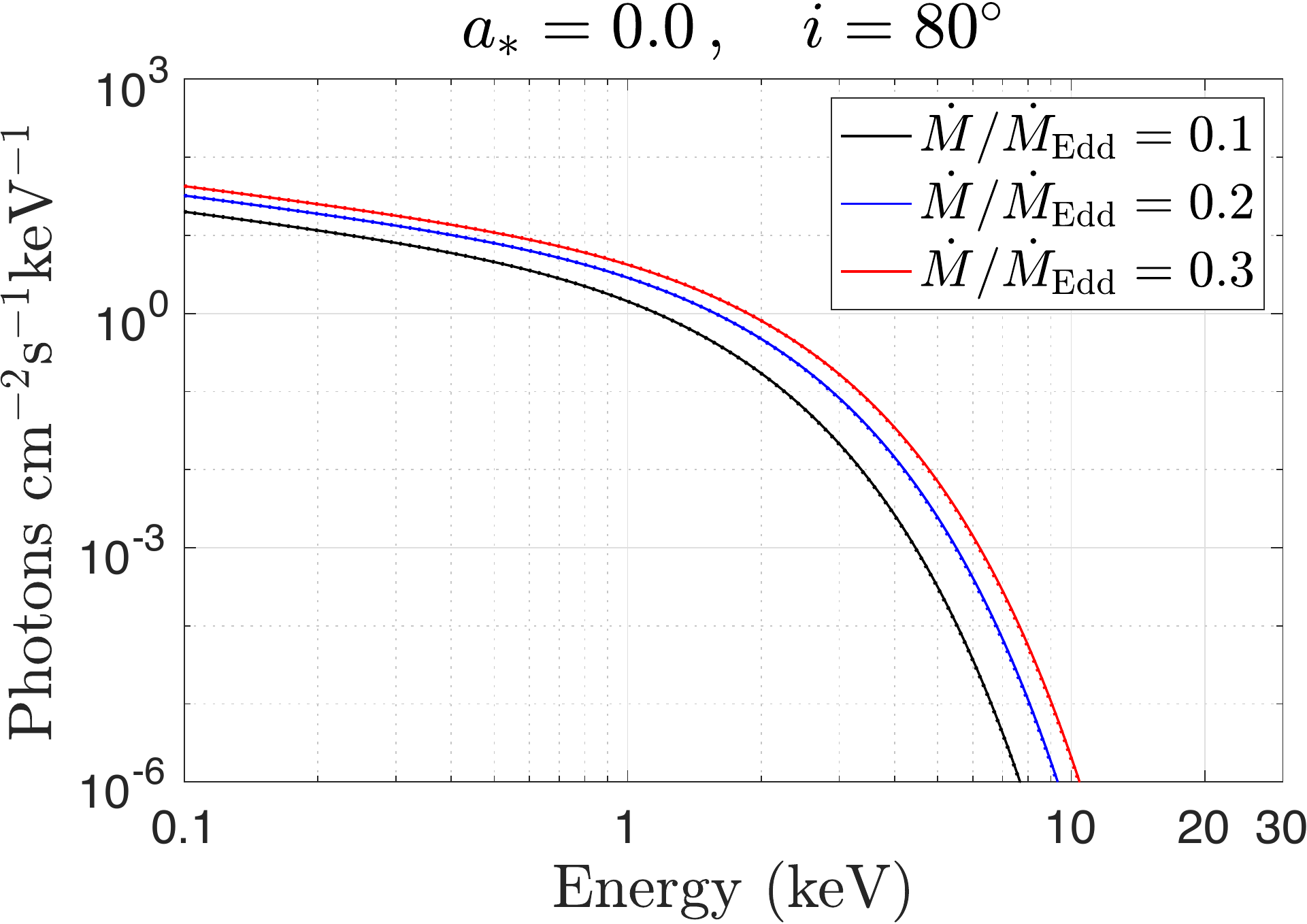} \\ \vspace{0.5cm}
\includegraphics[width=0.3\textwidth]{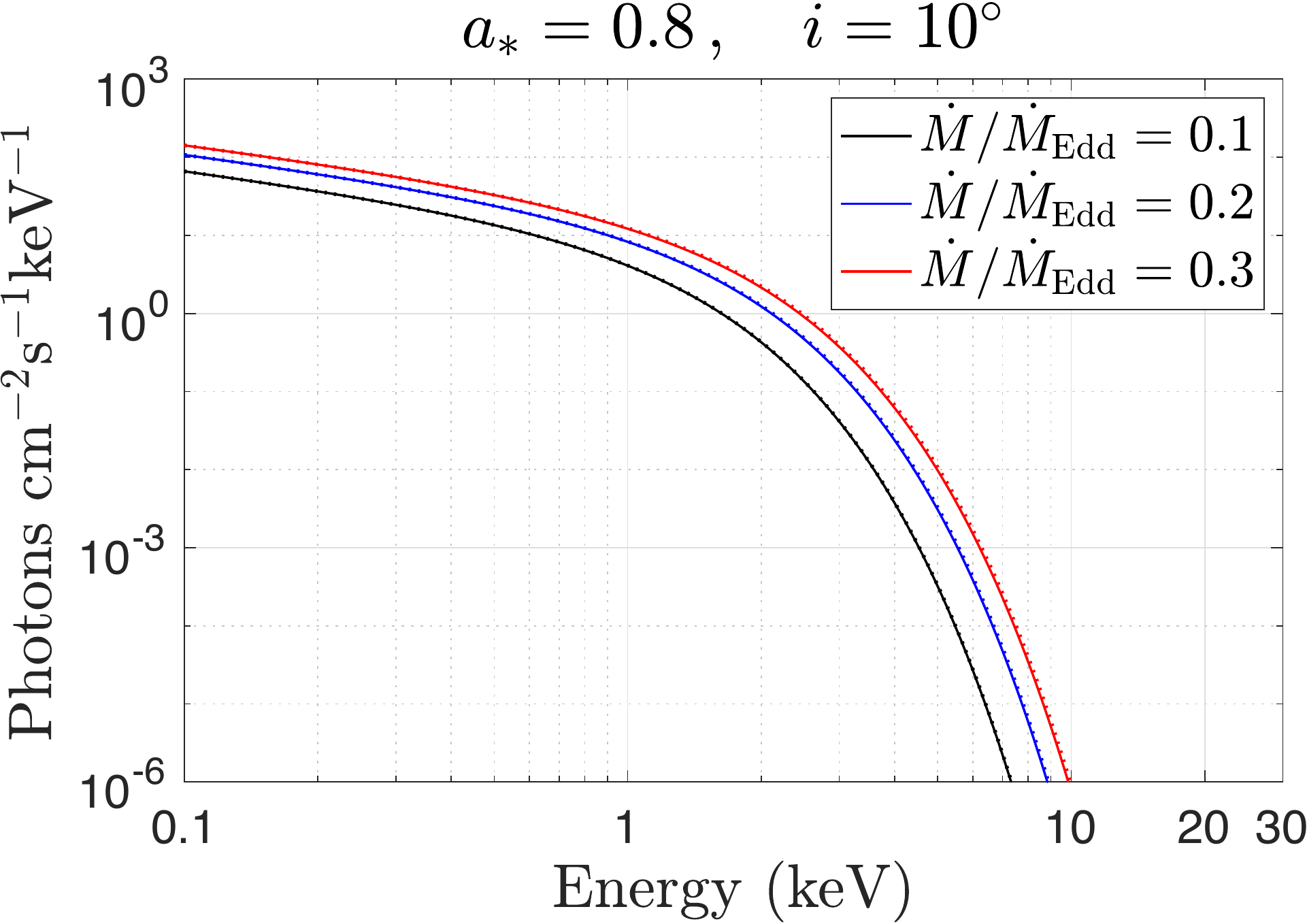} \hspace{0.4cm}
\includegraphics[width=0.3\textwidth]{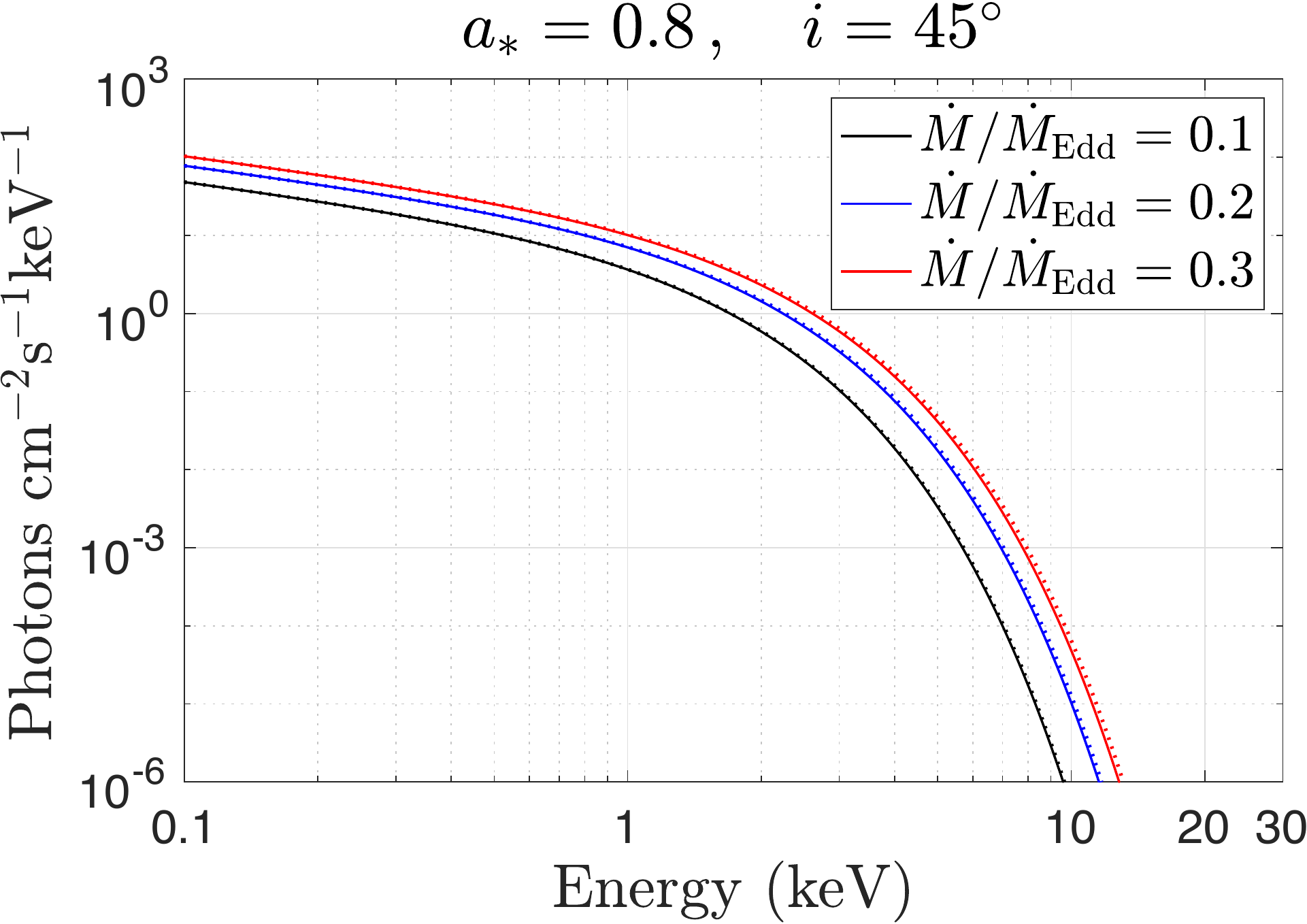} \hspace{0.4cm}
\includegraphics[width=0.3\textwidth]{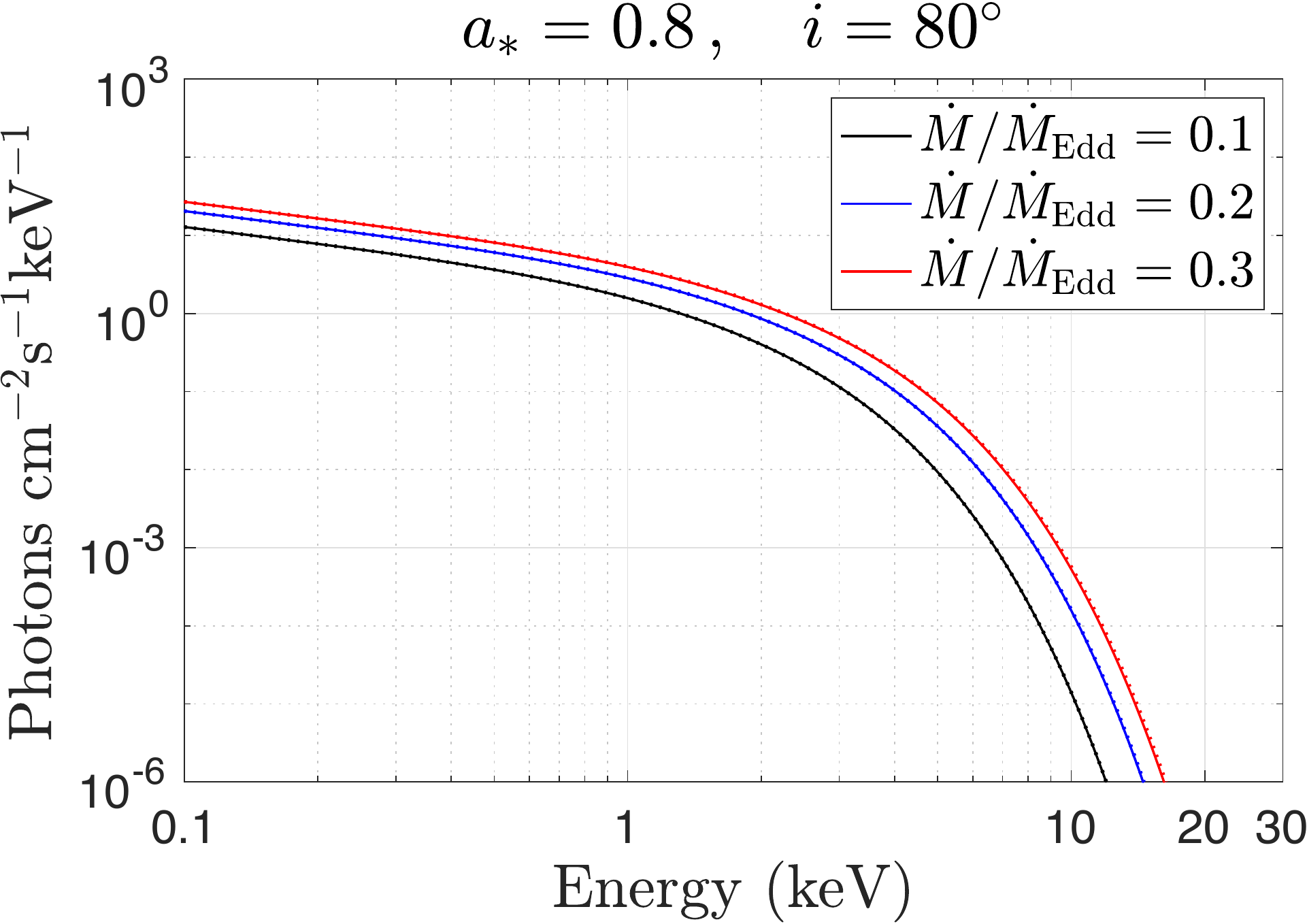} \\ \vspace{0.5cm}
\includegraphics[width=0.3\textwidth]{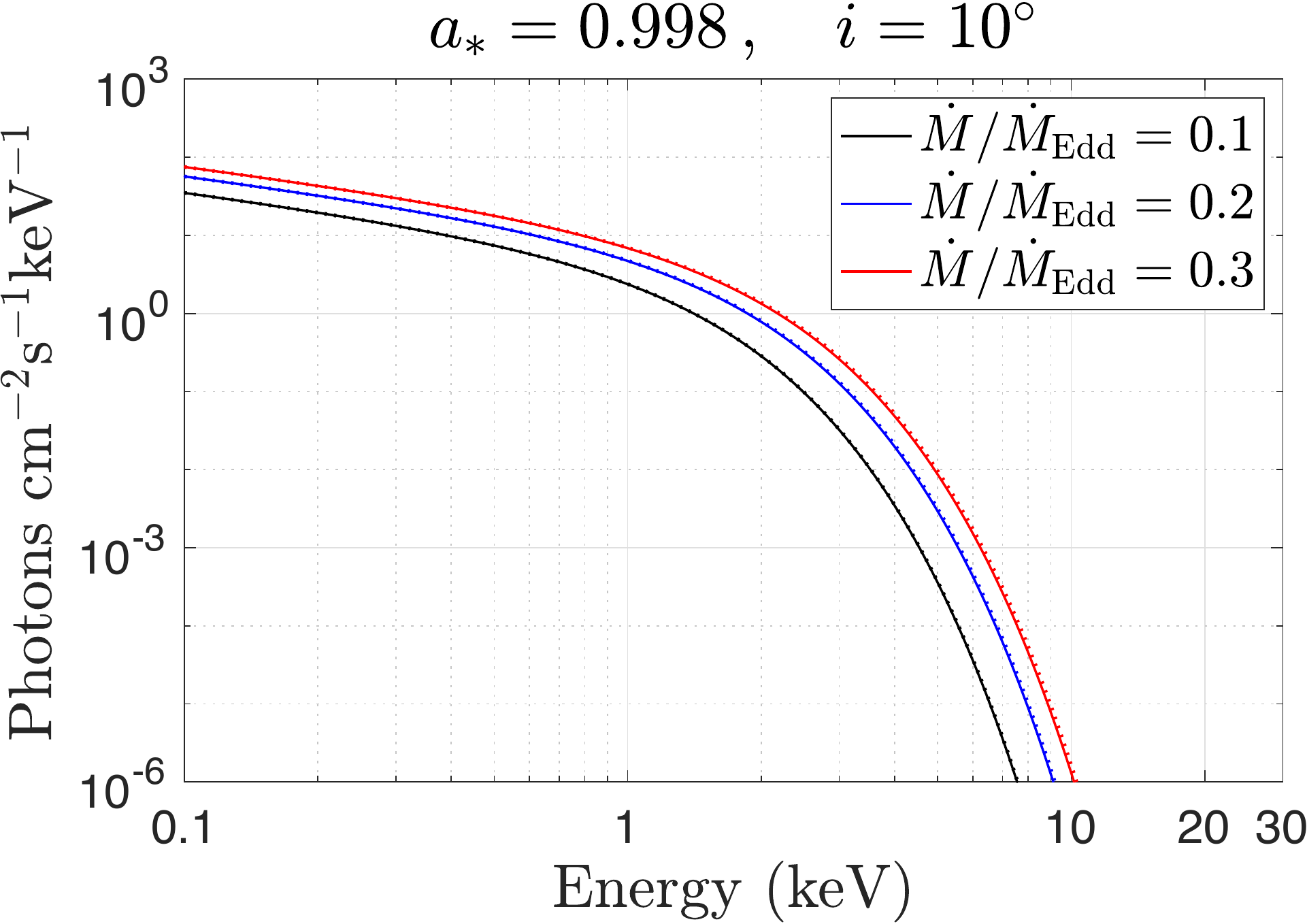} \hspace{0.4cm}
\includegraphics[width=0.3\textwidth]{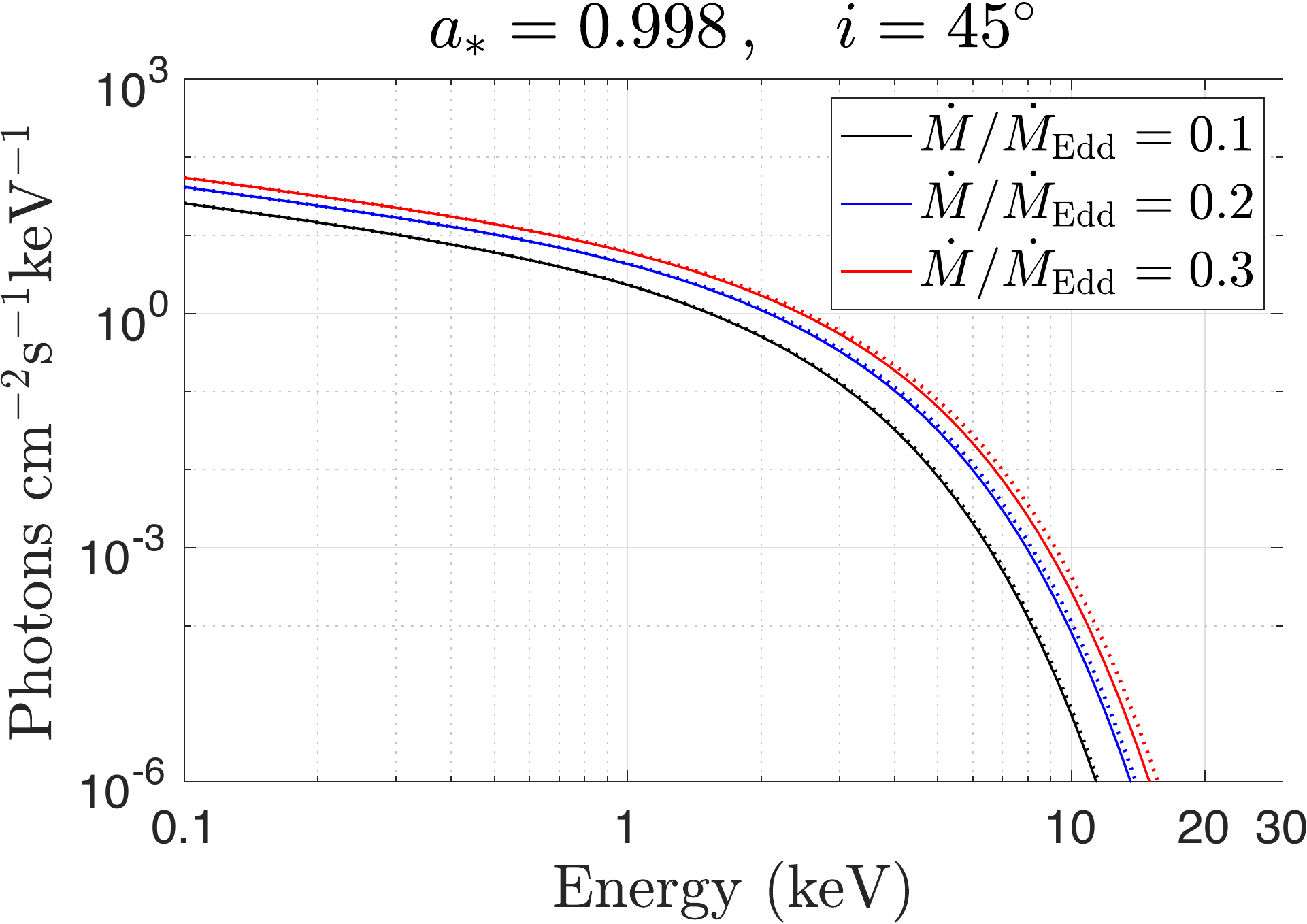} \hspace{0.4cm}
\includegraphics[width=0.3\textwidth]{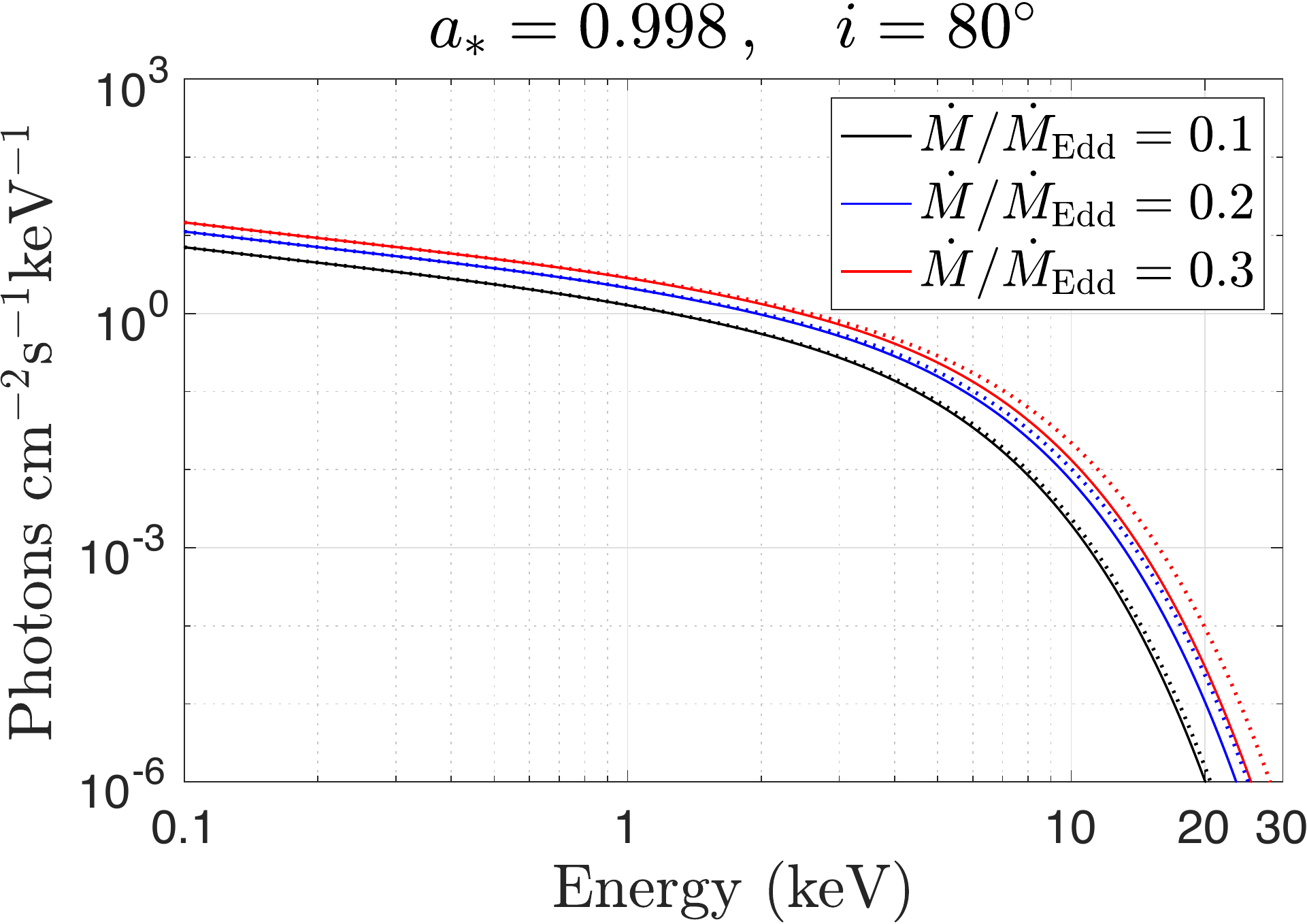}
\end{center}
\vspace{-0.1cm}
\caption{Synthetic thermal spectra for $a_* = 0$ (top row), 0.8 (central row), and 0.998 (bottom row), viewing angle $i = 10^\circ$ (left column), $45^\circ$ (central column), and $80^\circ$ (right column), and Eddington-scaled mass accretion rate $\dot{M} / \dot{M}_{\rm Edd} = 0.1$ (black curves), 0.2 (blue curves), and 0.3 (red curves). Solid curves are used for the thermal spectra for disks with finite thickness. Dotted curves are used for the thermal spectra for infinitesimally thin disks. The other parameters are: black hole mass $M = 10$~$M_\odot$, black hole distance $D = 10$~kpc, color factor $f_{\rm col} = 1$. \label{fig:spectra}}
\end{figure*}

The specific intensity of the radiation at the emission point is
\be
I_{\rm e} (\nu_{\rm e}) = \frac{2 h \nu_{\rm e}^3}{f_{\rm col}^4} \frac{\Upsilon}{\exp \left(\frac{h \nu_{\rm e}}{k_{\rm B} T_{\rm col}}\right) - 1} \, ,
\ee
where $h$ is Planck's constant, $k_{\rm B}$ is the Boltzmann constant, $\Upsilon$ is a possible parameter of order 1 that depends on the angle between the normal to the disk and the propagation direction of the photon (but in what follows we will assume $\Upsilon = 1$, corresponding to isotropic emission), and $f_{\rm col}$ is the color factor (a phenomenological parameter to take non-thermal effects into account, mainly the electron scattering in the disk atmosphere). $T_{\rm col} = f_{\rm col} T_{\rm eff}$ is the color temperature, and $T_{\rm eff}$ is the effective temperature of the accretion disk, which is obtained assuming $\mathcal{F} = \sigma T_{\rm eff}^4$, where $\sigma$ is the Stefan-Boltzmann constant and $\mathcal{F}$ the time-averaged energy flux emitted from the disk surface~\citep{2017bhlt.book.....B,2012ApJ...761..174B}
\be\label{eq-ff}
\mathcal{F} = \frac{\dot{M}}{4 \pi M^2} F(r) \, .
\ee
$F(r)$ is a dimensionless function that depends on the spacetime geometry only. For an infinitesimally thin disk, the mass accretion rate enters only via Eq.~(\ref{eq-ff}). The temperature profile is the same for infinitesimally thin disk and finite thickness disk.

The transfer function encodes the details of the spacetime metric and of the disk geometry. The transfer function is calculated with a ray-tracing code, firing photons from the screen of the distant observer backward in time to the accretion disk. The numerical scheme has been already discussed in~\citet{2017ApJ...842...76B}, \citet{2019ApJ...878...91A}, and \citet{2019PhRvD..99j4031Z}. The transfer functions are tabulated and stored into a FITS file. The main grid of the FITS file is 3-dimensional, for the black hole spin parameter $a_*$, the Eddington-scaled mass accretion rate $\dot{M}/\dot{M}_{\rm Edd}$, and the inclination angle of the disk $i$. The grid is $30 \times 30 \times 22$, namely we have 30~values for $a_*$ and $\dot{M}/\dot{M}_{\rm Edd}$ and 22~values for $i$. $a_*$ and $i$ have the same spacing as in our previous version of {\sc nkbb}~\citep{2019PhRvD..99j4031Z}. The values of $\dot{M}/\dot{M}_{\rm Edd}$ are evenly distributed over the range 0 to 0.3. Fig.~\ref{fig:grid} shows the grid points of the FITS file on the plane spin parameter $a_*$ vs mass accretion rate $\dot{M}/\dot{M}_{\rm Edd}$. For every set of $(a_*,\dot{M}/\dot{M}_{\rm Edd}, i)$, the transfer function is evaluated at 100~emission radii $r_{\rm e}$, from the ISCO to $10^6$~$M$. For every emission radius, the transfer function is evaluated at 40~values of $g^*$.

Once the transfer function is stored in the FITS file for some specific accretion disk model, we can calculated the thermal spectrum of the accretion disk using Eq.~(\ref{eq-F}). Fig.~\ref{fig:spectra} shows the output of our model for three different spin parameters ($a_* = 0$, 0.8, and 0.998, from top to bottom), three different inclination angles of the disk ($i=10^\circ$, $45^\circ$, and $80^\circ$, from left to right), and three different mass accretion rates ($\dot{M} / \dot{M}_{\rm Edd} = 0.1$, 0.2, and 0.3, respectively black, blue, and red curves). The dotted curves are for infinitesimally thin accretion disks and the solid curves are for finite thickness disks. In some cases, the difference between the spectra of infinitesimally thin disks and finite thickness disks is hard to see. The general trend is that the discrepancy between the two spectra increases for higher values of the spin parameter and the inclination angle. The spectra in Fig.~\ref{fig:spectra} are calculated including self-shadowing of the inner part of the disk~\citep{2018ApJ...855..120T,2018ApJ...868..109T}. The effect of self-shadowing is to obscure a part of the inner disk to the distant observer, thus reducing the total photon count; the impact on the thermal spectrum studied here is much weaker than the impact on the reflection spectrum studied in~\citet{2018ApJ...855..120T}, mainly because the emission profile of the former is less steep than that of the latter.


\section{Impact of the disk thickness on the spin measurement: GRS~1915+105}\label{s-grs}

In order to evaluate the impact of the thickness of the disk on the estimate of black hole spins, we analyze an X-ray observation of a black hole and we measure its spin parameter with {\sc nkbb}, either with the standard assumption of an infinitesimally thin disk (model~1 in this and next section) and with the disk model with finite thickness (model~2 in this and next section). The difference between the two spin measurements provides an estimate of the systematic uncertainty due to the disk thickness in the models with infinitesimally thin disks. From Fig.~\ref{fig:spectra} we expect that the difference between the spectra of the two disk models is larger when the source has its spin parameter $a_*$ close to 1 and the inclination angle of the accretion disk $i$ is high. We thus decide to test the new model with the black hole binary GRS~1915+105.

GRS~1915+105 is a low-mass X-ray binary. The distance of the source is $D = 8.6_{-1.6}^{+2.0}$~kpc and the mass of its black hole is $M = 12.4_{-1.8}^{+2.0}$~$M_\odot$~\citep{2014ApJ...796....2R}. Employing models with infinitesimally thin accretion disks, the black hole spin parameter has been estimated by several authors and different observations, always finding a value close to 1. \citet{2006ApJ...652..518M} find $a_* > 0.98$ with the continuum-fitting method and the analysis of \textsl{RXTE} and \textsl{ASCA} data. \citet{2009ApJ...706...60B} and \citet{2013ApJ...775L..45M} analyze, respectively, a \textsl{Suzaku} and a \textsl{NuSTAR} observation of GRS~1915+105; both studies find $a_* = 0.98 \pm 0.01$ (1-$\sigma$ statistical error) from the analysis of the refection spectrum of the source. \citet{2019ApJ...884..147Z} and \citet{2020arXiv200309663A} reanalyze the \textsl{Suzaku} data of \citet{2009ApJ...706...60B} without assuming the Kerr metric and find that the black hole spin parameter is very close to 1 even in the presence of possible deviations from the Kerr geometry. The inclination angle of the accretion disk is also thought to be high, even if its exact value is a bit controversial. Assuming that the jet of the source is parallel to the black hole spin and orthogonal to the accretion disk, the inclination angle is $i = 66^\circ \pm 2^\circ$~\citep{1999MNRAS.304..865F}. From the analysis of the reflection spectrum of the accretion disk, one can find an independent estimate of the inclination angle of the inner part of the disk, with $i$ ranging from $\sim 60^\circ$ up to $\sim 80^\circ$~\citep{2009ApJ...706...60B,2013ApJ...775L..45M,2019ApJ...884..147Z,2020arXiv200309663A}.

The continuum-fitting method requires thermal dominant spectral data, which are defined by three conditions in~\citet{rem}: $i)$ the flux of the thermal component accounts for more than 75\% of the total 2-20~keV unabsorbed flux, $ii)$ the root mean square (RMS) variability in the power density spectrum in the 0.1-10~Hz range is lower than 0.075, and $iii)$ quasi-periodic oscillations (QPOs) are absent or very weak. Imposing these conditions, in the \textsl{RXTE} archive \citet{2006ApJ...652..518M} find 20~observations of GRS~1915+105, which become 5~observations after requiring that the Eddington-scaled luminosity is less than 30\%. Since our goal here is only to illustrate the impact of the disk thickness, and not to repeat a detailed measurement of the black hole spin of GRS~1915+105, we consider only one of these observations, corresponding to observation number~20 in \citet{2006ApJ...652..518M}. The observation was on 24~November 2003 and the exposure time was around 4.2~ks.

For the data reduction, we follow the standard data reduction approach, extracting PCA spectral in the \texttt{standard2f} mode and discarding data within 10 minutes from the South Atlantic Anomaly (SAA). We use the spectra of PCU-2 only because it is the best calibrated of the five PCU units and has been on during all the analyzed observations. We improve the calibration using the {\sc pcacorr}~tool \citep{Garcia:2014cra}; using this approach, we are able to reduce the systematic error applied to the data to only 0.1\%, applied to all PCU~2 energy channels.

The data are fitted with the XSPEC model~\citep{2006ApJ...652..518M} 

\vspace{0.1cm}
{\sc tbabs$\times$smedge$\times$gabs$\times$(nkbb + powerlaw)} .
\vspace{0.1cm}

{\sc tbabs} describes the Galactic absorption~\citep{2000ApJ...542..914W} and we freeze the hydrogen column density to $N_{\rm H} = 8 \cdot 10^{22}$~cm$^{-2}$~\citep{2020arXiv200309663A}; however, its exact value does not appreciably affect the fit because the \textsl{RXTE} data do not cover low energies. {\sc smedge} describes a broad iron absorption edge~\citep{1994PASJ...46..375E} and {\sc gabs} is a Gaussian absorption line around 7~keV. {\sc nkbb} describes the thermal spectrum of the accretion disk; we first consider the case of an infinitesimally thin disk (model~1) and then the case of a disk with finite thickness (model~2). We freeze the black hole mass to $M = 12.4$~$M_\odot$ and the black hole distance to $D = 8.6$~kpc \citep{2014ApJ...796....2R}. The inclination angle of the disk is frozen to $i = 73^\circ$, which is the measurement obtained from X-ray reflection spectroscopy in \citet{2020arXiv200309663A}; however, even in this case the exact value does not have a significant impact on the fit. We ignore the uncertainties of $M$, $D$, and $i$ as an accurate spin measurement of the black hole in GRS~1915+105 is beyond the scope of our study, which is only focused on a preliminary estimate of the impact of the disk thickness on spin measurements. {\sc powerlaw} describes a power law component.

The best-fit values of the two models are reported in Tab.~\ref{t-fit}, where all parameter uncertainties are at the 90\% of confidence level for one relevant parameter. As we can see, the best-fit values of the two models are all consistent. While we find quite high best-fit values for the photon index $\Gamma$, our measurements are consistent with that reported in \citet{2006ApJ...652..518M}. The spin measurements of the two models are
\be
0.9859 < a_* < 0.9903 \,\,\, \text{(model~1)} \\
0.9899 < a_* < 0.9962 \,\,\, \text{(model~2)}
\ee 
Note that here we are only considering the statistical uncertainty of the fit. We are ignoring all systematic uncertainties of the model and the contribution from the uncertainties on $M$, $D$, and $i$, three quantities that are usually poorly constrained and tend to dominate the final error on the black hole spin parameter~\citep{2011MNRAS.414.1183K,2014SSRv..183..295M}

\begin{table}
\centering
{\renewcommand{\arraystretch}{1.3}
\begin{tabular}{lcc}
\hline\hline
Model & 1 & 2 \\
\hline
{\sc tbabs} &&\\
$n_{\rm H}$ [$10^{22}$~cm$^{-2}$] & 8$^\star$ & 8$^\star$ \\
\hline
{\sc smedge} &&\\
$E_{\rm s}$ [keV] & $7.62_{-0.06}^{+0.06}$ & $7.62_{-0.08}^{+0.09}$ \\
$\tau_{\rm s}$ & $0.97_{-0.12}^{+0.24}$ & $0.95_{-0.17}^{+0.16}$ \\
\hline
{\sc gabs} &&\\
$E_{\rm line}$ [keV] & $7.07_{-0.05}^{+0.05}$ & $7.07_{-0.06}^{+0.07}$ \\
$\sigma$ [keV] & 0.5$^\star$ & 0.5$^\star$ \\
\hline
{\sc nkbb} &&\\
$M$ [$M_\odot$] & 12.4$^\star$ & 12.4$^\star$ \\
$D$ [kpc] & 8.6$^\star$ & 8.6$^\star$ \\
$i$ [deg] & 73$^\star$ & 73$^\star$ \\
$a_*$ & $0.9881_{-0.0022}^{+0.0022}$ & $0.9926_{-0.0027}^{+0.0036}$ \\
$\dot{M}$ [$\dot{M}_{\rm Edd}$] & $0.183_{-0.005}^{+0.003}$ & $0.1834_{-0.0011}^{+0.0017}$ \\
$f_{\rm col}$ & 1.7$^\star$ & 1.7$^\star$ \\
\hline
{\sc powerlaw} &&\\
$\Gamma$ & $3.78_{-0.03}^{+0.03}$ & $3.78_{-0.05}^{+0.05}$ \\
norm & $62_{-3}^{+4}$ & $63_{-5}^{+5}$ \\
\hline
$\chi^2/\nu$ & $54.97/39=1.410$ & $55.18/39=1.415$ \\
\hline\hline
\end{tabular}}
\caption{Summary of the best-fit values for model~1 (infinitesimally thin disk) and model~2 (disk with finite thickness). The reported uncertainties correspond to the 90\% of the confidence level for one relevant parameter. $^\star$ indicates that the parameter is frozen in the fit. \label{t-fit}}
\end{table}


\section{Discussion}\label{s-d}

In the traditional framework of the continuum-fitting method, the model depends on 5~parameters: the black hole mass $M$, the black hole spin parameter $a_*$, the black hole distance $D$, the mass accretion rate $\dot{M}$, and the inclination angle of the disk $i$\footnote{The color factor $f_{\rm col}$ can be calculated with a model for the disk atmosphere and mainly depends on the mass accretion rate $\dot{M}$, so it is not a free parameter~\citep{2014SSRv..183..295M}.}. However, the resulting spectrum is simply a multi-temperature blackbody-like spectrum without particular features and it is not possible to infer the values of all the free parameters from the fit. It is thus necessary to have independent estimates of the black hole mass, black hole distance, and inclination angle, often obtained from optical observations, and then one can fit the thermal component of the source to measure the black hole spin parameter and the mass accretion rate~\citep{1997ApJ...482L.155Z,2011CQGra..28k4009M,2014SSRv..183..295M}.

The model commonly used for the continuum-fitting method is {\sc kerrbb}~\citep{2005ApJS..157..335L}, and its extension {\sc kerrbb2}. There are now about 15~stellar-mass black holes with an estimate of the spin parameter via the continuum-fitting method. The model employs Novikov-Thorne disks and assumes that the disks are infinitesimally thin. The impact of the theoretical model of {\sc kerrbb} on the estimate of black hole spins has been investigated in \citet{2010MNRAS.408..752P} and \citet{2011MNRAS.414.1183K}: the authors ran GRMHD simulations of thin accretion disks for different values of the black hole spin parameter, calculated the spectra emitted by their simulated disks, evaluated the differences with the spectra calculated from the Novikov-Thorne model with an infinitesimally thin disk and no emission inside the ISCO, and eventually estimated the errors on the spin measurements obtained with an infinitesimally thin disk model. The conclusion of \citet{2011MNRAS.414.1183K} is that the uncertainties on current spin measurements with the continuum-fitting method are dominated by observational uncertainties, while systematic uncertainties due to the theoretical model are negligible. Their conclusion is consistent with our results, where the estimates of the black hole spin with models~1 and 2 overlap at 90\% of confidence level and we have not included the larger uncertainty contributions from the errors on $M$, $D$, and $i$.

An extension of {\sc kerrbb}/{\sc kerrbb2} was presented in~\citet{2011A&A...533A..67S} and called {\sc slimbb}, as capable of describing the thermal spectra of thin and slim accretion disks~\citep{2011A&A...527A..17S}, so valid up to Eddington-scaled luminosities $\sim 0.7$. {\sc slimbb} not only takes the thickness of the disk into account, but includes also the radial advection of heat, which changes the emission profile, and deviations of the inner edge of the disk from the ISCO radius, both effects important at high mass accretion rates. \citet{2011A&A...533A..67S} analyze a large number of \textsl{RXTE} data of the black hole binary LMC~X-3, which is thought to be a black hole with a moderate value of the spin parameter ($a_* < 0.7$) and a high disk inclination angle ($i \sim 70^\circ$). \citet{2011A&A...533A..67S} find no discrepancy between the black hole spin measurements obtained with {\sc kerrbb}/{\sc kerrbb2} and {\sc slimbb} when the Eddington-scaled accretion luminosity of the source is below 30\%.

The impact of the disk structure has been investigated even for spin measurements obtained from the analysis of the reflection spectrum of the disk. X-ray reflection spectroscopy can potentially provide more precise spin measurements, because the latter do not require independent estimate of the black hole mass, distance, and disk inclination angle, three quantities that are often difficult to measure and are affected by large systematic uncertainties. Moreover, the reflection spectrum has more features than the thermal component, and this also helps constraining the model parameters. However, the model itself is more complicated and the theoretical uncertainties in the model can have a larger impact on the final spin measurement.

\citet{2008ApJ...675.1048R} simulate geometrically thin accretion disks in a pseudo-Newtonian potential. They find that spin measurements employing the standard infinitesimally thin disk model lead to overestimate the black hole spin, which is the result contrary to ours, but the main contribution to the final spin measurement is not determined by the disk thickness but by the radiation emitted from the plunging region inside the ISCO.

In our work, we have implemented the disk model proposed in \citet{2018ApJ...855..120T}, where the authors study the impact of the disk thickness on X-ray reflection spectroscopy spin measurements. \citet{2018ApJ...855..120T} find that the analysis with an infinitesimally thin disk model leads to underestimate the black hole spin, like in our analysis for the continuum-fitting method, but it should be noted that the coronal geometry and the corresponding intensity profile play quite an important role on the actual impact of the disk geometry, so a direct comparison is not straightforward. The systematic uncertainty on the final spin measurement from an infinitesimally thin disk is not negligible in \citet{2018ApJ...855..120T}. \citet{2020arXiv200309663A} employ the disk geometry of \citet{2018ApJ...855..120T} and analyze \textsl{Suzaku} data of the black hole binary GRS~1915+105, where the coronal geometry is presumably different from the point-like lamppost source of \citet{2018ApJ...855..120T}, and they find no appreciable different between the spin measurement of a reflection spectrum that assumes an infinitesimally thin disk and a disk of finite thickness.

The thickness of thin disks has surely some impact on the final spin measurement of an accreting black hole, either we use the continuum-fitting method or X-ray reflection spectroscopy. For the continuum-fitting method, where there is not the problem of the coronal geometry as in the analysis of the reflection spectrum, it is straightforward to arrive at a conclusion. At present, the precision of spin measurements are limited by the observational measurements rather than by the theoretical model, as already pointed out in \citet{2011MNRAS.414.1183K}. However, future observational measurements will be likely more precise and accurate, and in such a case the thickness of thin disks may become a new ingredient to be included in the theoretical model. Moreover, the impact of the thickness of the disk increases as the inclination angle increases. For higher inclination angles and/or thicker disks than those in the \textsl{Suzaku} observation of GRS~1915+105, a more significant part of the very innermost region of the accretion disk may be obscured by the disk itself~\citep{2018ApJ...855..120T}, and this could increase the difference in the best-fit values from the models with infinitesimally thin disk and disk with finite thickness.


\section*{Acknowledgements}

We wish to thank Jiachen Jiang for an earlier collaboration on the subject of this paper and useful discussions and suggestions.
This work was supported by the Innovation Program of the Shanghai Municipal Education Commission, Grant No.~2019-01-07-00-07-E00035, and the National Natural Science Foundation of China (NSFC), Grant No.~11973019.
V.G. is supported through the Margarete von Wrangell fellowship by the ESF and the Ministry of Science, Research and the Arts Baden-W\"urttemberg.
C.B., V.G., and A.T. are members of the International Team~458 at the International Space Science Institute (ISSI), Bern, Switzerland, and acknowledge support from ISSI during the meetings in Bern.








\bsp	
\label{lastpage}
\end{document}